\documentclass[aps,prl,twocolumn,groupedaddress]{revtex4}
\usepackage{graphicx}
\begin{document}

\title{Kafri, Mukamel, and Peliti Reply}

{\bf Kafri, Mukamel, and Peliti Reply}: The Comment of Hanke and
Metzler \cite{HM} questions the validity of the analysis presented
in \cite{KMP1} to DNA chains of finite length as used in
experiments. Their argument is that for the analysis to be valid
``each of the three segments going out of a vertex must be much
longer than the persistence length $\ell_p$ of this segment''. By
using the persistence lengths $\ell_p(L) \sim 40$\AA~for a single
strand and $\ell_p(H) \sim 500$ \AA~for a double helix (bound
segment) they arrive at the conclusion that in order to observe
the asymptotic behavior found in \cite{KMP1} one needs chains
which are far longer than those studies experimentally.

This assertion constitutes a misunderstanding of the analysis
given in \cite{KMP1}. In this analysis one considers a loop
interacting with {\it the rest of the chain} and not just with the
vicinal double helices. Thus, in \cite{KMP1} each of the two lines
attached to the loop is in fact composed of an alternating
sequence of bound segments and denaturated loops. It may be viewed
as a stick and joint structure, whereby adjacent double helices
(which may be considered as rigid rods as they are shorter than
$\ell_p(H)$) are loosely attached to each other via an open loop
(see Fig. 1). Thus, the ``rest of the chain'' as considered in
reference \cite{KMP1} is in fact a random rod structure with a
persistence length given by the length of the rod connecting two
loops. This length can be easily estimated from the analysis of
Poland and Scheraga and its generalization as given in
\cite{KMP1}. It can be shown that the probability distribution of
rods of length $k$ at the transition is $P(k) \sim (w/s)^k$. Here
$w=\exp(\beta_{\rm M} E_0)$ is the Boltzmann weight of a bound
pair with energy $-E_0$ at the melting temperature, $1/ \beta_{\rm
M}$, and $s$ is a non-universal geometrical factor. This is an
exponentially decaying distribution with a typical length of order
$\xi \sim 1/ \vert \ln(w/s) \vert$. This length has nothing to do
with the persistence length of the bounded segments and in fact it
is far shorter (of the order of a small number of base pairs). An
estimate of this length can be obtained using Eq. 4 of \cite{KMP1}
at the transition
\begin{equation}
\frac{s}{w}=\sum_{k=1}^{\infty} \frac{1}{k^c}+1.
\end{equation}
With the estimated value of $c=2.115$ (in three dimensions) this
yields $\xi \simeq 1.07$ in units of $\ell_p(L)$. Note that this
length is non-universal and it depends on the details of the
model, such as the statistical weight of small loops, the stacking
interactions and other details. However, these features are not
expected to change this length considerably. Therefore, the
criticism expressed in the Comment is not valid and the analysis
in the Letter is applicable to finite chains of the length studied
experimentally. In reference \cite{KMP1} the detailed structure of
the ``rest of the chain'' is ignored when the self-avoiding
interaction is considered. However, as argued in \cite{KMP1} this
assumption is reasonable and should yield a good estimate of the
effect of self-avoiding interactions.

\begin{figure}
\includegraphics[width=1.5cm,angle=90]{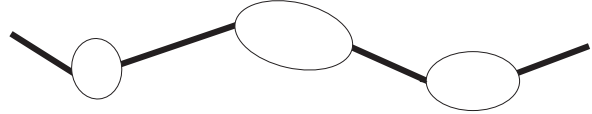} \caption{Schematic
representation of a microscopic configuration of the DNA molecule.
Here the flexibility of the ``{\it rest of the chain}'' emerges
from the presence of loops which connect short but more rigid
bound segments.}
\end{figure}

A strong support for this picture is provided by numerical
simulations of Causo {\it et. al.} \cite{CCG} and Carlon {\it et.
al.} \cite{COS} of chains of finite length in which the self
avoiding interactions are fully taken into account. In both
studies the melting transition was found to be first order,
compatible with $c>2$ as found in \cite{KMP1}. In \cite{COS} the
exponent $c$ is evaluated directly for chains of length 50-200
monomers (each monomer corresponds to $\ell_p(L)$ namely to about
8 base pairs). Moreover, the inclusion of the different
stiffnesses of the bound and unbound segments (in simulations
persistence length ratio, $\ell_p(H) /\ell_p(L)$, was taken to
vary between 1 and 10) did not change the estimate significantly.
The fact that the observed $c$ is consistent with that obtained by
reference \cite{KMP1} strongly supports the basic assumptions
behind this analysis. It also indicates that the scaling behavior
predicted in \cite{KMP1} may be observed in finite chains of
lengths between hundreds to thousands base pairs, which is the
experimentally relevant range.

\noindent Y. Kafri and D. Mukamel

Department of Physics of Complex Systems, The Weizmann Institute
of Science, Rehovot 76100, Israel

\noindent L. Peliti

Dipartimento di Scienze Fisiche and Unit\`a INFM, Universit\`a
``Federico II'', Complesso Monte S. Angelo, I--80126 Napoli, Italy

\end{document}